\documentclass{jpp}
\usepackage{graphicx}
\usepackage{epstopdf, epsfig}
\usepackage{color}

\expandafter\let\csname equation*\endcsname\relax
\expandafter\let\csname endequation*\endcsname\relax
\usepackage{amsmath}

\newcommand{\beq}{\begin{equation}}
\newcommand{\eeq}{\end{equation}}
\newcommand{\vths}[1]{\ensuremath{v_{\mathrm{th},#1}}}

\newcommand{\dphi}{\varphi}
\newcommand{\pd}[2]{\frac{\partial #1}{\partial #2}}
\newcommand{\gyroR}[1]{\left<#1\right>_{\mathbf{R}_{s}}}
\newcommand{\hks}{\hat{h}_{\mathbf{k},s}}
\newcommand{\mbf}[1]{\mathbf{#1}}

\newcommand{\vpar}{v_{\parallel}}
\newcommand{\kpar}{k_{\parallel}}
\newcommand{\iunit}{\textnormal{i}}
\newcommand{\real}[1]{\textnormal{Re}\left(#1\right)}
\newcommand{\imag}[1]{\textnormal{Im}\left(#1\right)}
\newcommand{\erfi}[1]{\textnormal{erfi}\left(#1\right)}
\newcommand{\erfc}[1]{\textnormal{erfc}\left(#1\right)}
\newcommand{\zi}{\textnormal{i}}

\newcommand{\pflx}{\Gamma}
\newcommand{\modgam}{\Upsilon}

\newcommand{\defeq}{\doteq}

\newcommand{\hswap}[1]{h_{#1}^{\leftrightarrow}}
\newcommand{\phiswap}{\dphi^{\leftrightarrow}}

\shorttitle{Turbulent heating in an inhomogeneous, magnetized plasma slab}
\shortauthor{M. Barnes, P. Abiuso and W. Dorland}

\title{Turbulent heating in an inhomogeneous, magnetized plasma slab}

\author{Michael Barnes\aff{1},\aff{2}
  \corresp{\email{michael.barnes@physics.ox.ac.uk}},
  P. Abiuso\aff{3}
  \and W. Dorland\aff{4}}

\affiliation{\aff{1}Rudolf Peierls Centre for Theoretical Physics, University of Oxford,
Oxford OX2 8ES, UK
\aff{2}Euratom/CCFE Fusion Association, Culham Science Centre, Abingdon OX14 3DB, UK
\aff{3}Dipartimento di Fisica dell'Universit\`a di Pisa, Scuole Normale Superiore, I-56126 Pisa, Italy
\aff{4}Department of Physics, University of Maryland, College Park, Maryland 20740, USA}

\begin{document}

\maketitle

\begin{abstract}

Observational evidence in space and astrophysical plasmas with long collisional mean free path suggests that more massive
charged particles may be preferentially heated.  One possible mechanism for this is the
turbulent cascade of energy from injection to dissipation scales, where the energy
is converted to heat.  Here we consider a simple system consisting of a magnetized
plasma slab of electrons and a single ion species with a cross-field density gradient.  
We show that such a system is subject to an electron
drift wave instability, known as the universal instability, which is stabilized only when the electron and ion thermal speeds
are equal.  For unequal thermal speeds, we find that the instability gives rise to turbulent energy exchange between ions and electrons that acts to equalize the thermal speeds.  Consequently,
this turbulent heating tends to equalize the component temperatures of pair plasmas and to heat ions to much higher temperatures than electrons for conventional mass-ratio plasmas.

\end{abstract}

\section{Introduction}

An interesting and fundamental question in plasma physics is how
thermal equilibrium is determined in an essentially collisionless plasma.  In such a system, 
there is no reason \textit{a priori} to assume that the relative temperatures of the 
component species will be equal:  On the contrary, there are numerous collisionless
heating mechanisms that have been identified - shocks, magnetic reconnection,
cyclotron resonance~\citep{cranmerApJ99}, and various forms of turbulent heating~\citep{quataertApJ99a,howesJGR08,chandranApJ10a,barnesPRL12}, to name a few - and
each of these mechanisms may drive temperature separation instead of equilibration.  
Indeed, evidence from observations of collisionless space and astrophysical plasmas, e.g.,
the solar wind and accretion flows onto compact objects,
suggests that more massive charged particles may be preferentially heated; cf.
~\citep{schmidtGRL80,collierGRL96,kohlSP97,quataertApJ99a}.

A thorough treatment of this problem would require exploration of a parameter space
spanning a wide range of plasma $\beta$, electron-ion temperature ratio, energy injection 
mechanisms, etc.,
while including all potentially relevant physics present from the macroscopic to microscopic
space-time scales.  As such a study is infeasible, we choose to focus on a single heating 
mechanism - turbulent Joule heating - in a simplified
system: a two-component plasma immersed in a straight, homogenous magnetic field 
with a cross-field density gradient.  This system supports electron drift waves and is 
known to be susceptible to the so-called `universal' instability~\citep{galeevJETP63,krallPoF65,landremanPRL15,helanderPoP15}.
The universal instability serves as an energy injection mechanism for turbulence at the ion
Larmor scale that gives rise to plasma fluctuations that drive cross-field
transport and Joule heating (or cooling)
of the component plasma species.  For sufficiently infrequent collisions, this turbulent
heating and transport determines the thermal equilibrium of the plasma.

To guide our study, we note that the universal instability is derived analytically in the limit of disparate ion and electron thermal speeds.  
As noted in Ref.~\citep{helanderPRL14}, the universal instability is eliminated for an equal
temperature pair plasma, or, more generally, when the ion and electron thermal speeds are equal.  
In the following sections we analytically and numerically calculate the dependence of the 
universal instability and resultant turbulent heating on electron-ion temperature ratio
for two cases: pair plasmas and conventional mass ratio plasmas.  The former case is relevant
for proposed experiments in which
a small population of positrons is to be added to a pure electron plasma of a different temperature~\citep{pedersenPRL02};
the latter may provide insight into how astrophysical plasmas can achieve disparate electron and ion
temperatures.
We introduce the gyrokinetic model used for the analysis
in Sec. 2, and we calculate linear growth rates and quasilinear heating and cross-field flux
estimates in Sec. 3 before concluding.


\section{Model system}

We consider a collisionless plasma immersed in a straight, homogeneous magnetic field 
$\mathbf{B}=B\hat{\mathbf{z}}$
and with a fixed density gradient perpendicular to the field in the $x$-direction.  The fixed density
gradient could be, e.g., the result of gravitational equilibrium in an astrophysical plasma or of
an external particle source in a laboratory plasma.
We restrict our attention to electrostatic fluctuations whose frequency is small compared 
with the Larmor frequency and adopt the gyrokinetic ordering:
\beq
\epsilon \doteq \frac{\omega}{\Omega_s} \sim \frac{k_{\parallel}}{k_{\perp}} \sim \frac{\rho_s}{L_n} \sim \frac{\delta f_s}{f_s} \sim \frac{q_s\dphi}{T_s}\ll 1,
\eeq
where $\dphi$ is the electrostatic potential fluctuation, $f_s$ is the distribution function for species $s$, $\delta f$ is the fluctuating
component of $f$, $\omega$ is the characteristic frequency of the fluctuations, 
$\kpar$ and $k_{\perp}$ are the associated wavenumbers along and across the mean field, 
$\Omega_s$ is the Larmor frequency for species $s$, $\rho_s$ its thermal Larmor radius, 
$T_s$ its temperature, $q_s$ its charge, and $L_n$ is the mean density gradient scale
length.

Applying these orderings to the Fokker-Planck equation and averaging over the rapid gyration
of particles about the mean magnetic field results in the gyrokinetic equation,
\beq
\pd{h_s}{t} + \vpar\pd{h_s}{z} + \frac{c}{B}\{\gyroR{\dphi},h_s\} = -q_s \pd{\gyroR{\dphi}}{t}\pd{F_{0,s}}{E_s} + \frac{c}{B}\pd{\gyroR{\dphi}}{y}\pd{F_{0,s}}{x} + C[h_s],
\label{eqn:gke}
\eeq
where $h_s = \delta f_s - q_s\dphi (\partial F_{0,s}/\partial E_s)$ is the non-Boltzmann piece of $\delta f_s$, $E_s = m_s v^2 / 2$ is the kinetic energy of species $s$, $m_s$ is species mass,
$t$ is time, $\vpar$ is the parallel component of the particle velocity, $c$ is the speed of light,
$F_{0,s}$ is the mean component of $f_s$,
$\{\}$ indicates a Poisson bracket, $\gyroR{\dphi}$ is the
average of $\dphi$ over Larmor angle at fixed guiding center position $\mbf{R}_s$,
and $C[h_s]$ represents the effect of Coulomb collisions on species $s$.

The gyrokinetic system is closed by coupling to Poisson's equation:
\beq
4\pi\sum_s q_s \int d^3 v \delta f_s = -\nabla_{\perp}^2 \dphi,
\label{eqn:poisson}
\eeq
with $\nabla_{\perp}^2 = \partial_x^2 + \partial_y^2$.  If the 
Debye length $\lambda_s \doteq (T_s/4\pi n_s q_s^2)^{1/2}$ is much smaller than
the electron Larmor radius, the righthand side of Eq.~(\ref{eqn:poisson}) can be neglected.
In this limit Poisson's equation reduces to the quasineutrality constraint that the total 
charge density of the plasma is zero.  We note that a non-relativistic treatment of finite Debye
length effects requires a small plasma beta, $\beta_s=8\upi p_s/B^2 \lesssim (\vths{s}/c)^2$, consistent
with the electrostatic approximation we employ.

%

\subsection{Turbulent heating and heat transport}

By taking fluid moments of the Fokker-Planck equation and closing the system with the gyrokinetic
ordering, one obtains an equation for the slow
evolution of the mean temperature $T_{s}$:
\beq
\frac{3}{2}n_s\frac{dT_s}{dt} + \pd{Q_s}{x} = H_s,
\label{eqn:pressure}
\eeq
where the cross-field turbulent heat flux $Q_s$ and turbulent heating $H_s$ are given by
\beq
Q_s = \frac{1}{V}\int d^3 R\int_{\mbf{R}} d^3 v \left(\frac{m_s v^2}{2}-\frac{3}{2}T_s\right)\overline{\left(h_s v_E\right)}
\label{eqn:qflx}
\eeq
and
\beq
H_s = \frac{1}{V}\int d^3 R\int_{\mbf{R}} d^3 v q_s \overline{\left(h_s \pd{\gyroR{\dphi}}{t}\right)}.
\label{eqn:xchange0}
\eeq
Here $V$ is the volume of the region over which the spatial integration is performed,
the subscript $\mbf{R}$ on the velocity integration indicates that it is carried out at 
fixed guiding center, $v_E=-(c/B)\partial\dphi/\partial y$ is the x-component of the $E\times B$ drift
velocity, and the overline indicates an average over time scales long compared
to the fluctuation time $1/\omega$ but short compared to the equilibrium time scale.  
Note that we have neglected collisional temperature equilibration
as we are considering systems with collisional mean free path much longer than
any other scales of interest.

An alternative expression for $H_s$ is obtained by integrating Eq.~(\ref{eqn:xchange0}) by parts in 
time~\citep{candyPoP13}:
\beq
H_s = \frac{1}{2V}\int d^3 R\int_{\mbf{R}} d^3 v q_s \overline{\left(h_s \pd{\gyroR{\dphi}}{t}-\pd{h_s}{t} \gyroR{\dphi}\right)}.
\label{eqn:xchange}
\eeq
Substitution of Poisson's equation~(\ref{eqn:poisson}) in Eq.~(\ref{eqn:xchange})
immediately indicates that the net (species-summed) turbulent heating is zero in the absence of an external energy injection mechanism; i.e., for
a two-component plasma, $\overline{H_i} = -\overline{H_e}$.  If trace minority ions are present, one can obtain mass- and charge-dependent scalings for their turbulent heating rates relative to the main ions~\citep{barnesPRL12}.  Expressing Eqs.~(\ref{eqn:qflx}) and~(\ref{eqn:xchange}) in terms of Fourier modes with $h_s = \sum_{\mbf{k}} \tilde{h}_{\mbf{k},s}(\vpar,v_{\perp},t) \exp(\zi\mbf{k}\cdot\mbf{R}_s)$ gives
\beq
Q_s = \zi\frac{c}{B}\sum_{\mbf{k}} k_y\int d^3 v J_0(\alpha_{k_{\perp},s}) \frac{m_s v^2}{2}\overline{\left(\tilde{h}_{\mbf{k},s}  \tilde{\dphi}_{\mbf{k}}^*\right)} \doteq \sum_{\mbf{k}}\tilde{Q}_{\mbf{k},s}
\label{eqn:qflxk}
\eeq
and
\beq
H_s = \frac{q_s}{2}\sum_{\mbf{k}} \int d^3 v J_0(\alpha_{k_{\perp},s}) \overline{\left(\pd{\tilde{\dphi}_{\mbf{k}}^*}{t} \tilde{h}_{\mbf{k},s} - \tilde{\dphi}_{\mbf{k}}^* \pd{\tilde{h}_{\mbf{k},s}}{t}\right)} \doteq \sum_{\mbf{k}} \tilde{H}_{\mbf{k},s},
\label{eqn:xchangek}
\eeq
where $\alpha_{k_{\perp},s} \doteq k_{\perp} v_{\perp}/\Omega_s$ and $J_0$ is a Bessel function of the first kind.

\subsection{Connection to particle transport}

The effect of cross-field particle transport is encapsulated in the continuity equation:
\beq
\pd{n_s}{t} + \pd{\pflx_s}{x} = 0,
\label{eqn:continuity}
\eeq
where $n_s$ is particle density and
\beq
\pflx_s = \frac{1}{V}\int d^3 R\int_{\mbf{R}} d^3 v \overline{\left(h_s v_E\right)}
\label{eqn:pflx}
\eeq
is the cross-field particle flux.
Expanding $\dphi$ and $h$ in terms of Fourier modes gives
\beq
\pflx_s = \zi\frac{c}{B}\sum_{\mbf{k}} k_y\int d^3 v J_0(\alpha_{k_{\perp},s}) \overline{\left(\tilde{h}_{\mbf{k},s}  \tilde{\dphi}_{\mbf{k}}^*\right)} \doteq \sum_{\mbf{k}}\tilde{\pflx}_{\mbf{k},s}.
\label{eqn:pflxk}
\eeq

The particle flux $\pflx$ can be related to the turbulent heating of Eq.~(\ref{eqn:xchange0}) by
multiplying the gyrokinetic equation~(\ref{eqn:gke}) by $\dphi$ and averaging over the phase space:
\beq
H_s =  -\frac{1}{V} \int d^3 R\int_{\mbf{R}} d^3 v \left(\frac{T_s h_s}{F_{0,s}}C[h_s]\right) + T_{s} \pd{\ln n_{s}}{x} \pflx_s
\label{eqn:xchange2}
\eeq

Note that the effect of collisions is retained in Eq.~(\ref{eqn:xchange2}), but collisional temperature
equilibration is neglected in the energy equation~(\ref{eqn:pressure}).  This is because the turbulent
fluctuations undergo filamentation in phase space and thus tiny collisional deflections in velocity 
have immediate
impact on the turbulent fluctuations -- even when the amount of energy that is exchanged
in the course of the deflection is negligible.
At the small phase space scales where collisional dissipation occurs,
the dissipation is effectively diffusive.  This gives rise to
heating that is positive definite for each species.  Furthermore, 
Poisson's equation~(\ref{eqn:poisson}) applied to our two-component plasma dictates that $\pflx_i=\pflx_e \doteq \pflx$ and $n_{i}=n_{e}\doteq n$.  Substituting these expressions in Eq.~(\ref{eqn:xchange2}) and enforcing $H_i = -H_e$
results in the constraint that $\pflx (\partial \ln n/\partial x) \leq 0$ and thus
$|H_s| \leq T_s |\pflx (\partial \ln n/\partial x)|$.  Comparing this
inequality with Eqs.~(\ref{eqn:pressure}) and~(\ref{eqn:continuity}), one finds that the
timescale associated with turbulent particle transport is always at least as fast as that associated with 
turbulent heating.  One would therefore expect that when turbulent heating drives temperatures apart (as we find below), the effect would be bounded by the rate of transport down the density gradient.
Depending on the system under consideration, this may be set by the rate at which plasma is introduced.

\section{Linear analysis}

If we consider small amplitude perturbations, we may neglect the quadratic nonlinearity and
carry out a linear analysis of the gyrokinetic equation.  Upon assuming solutions of the form
$h_s = \sum_{\mbf{k}}\hks(\vpar, v_{\perp})\exp(i \mathbf{k}\cdot\mathbf{R}_s-i\omega t)$, one obtains
\beq
\hks = \frac{q_s \hat{\dphi}_{\mbf{k}}}{T_s}J_0(\alpha_{k_{\perp},s})\left( \frac{\omega + \omega_{*,s}}{\omega-\kpar \vpar}\right)F_{M,s},
\label{eqn:hks}
\eeq
where 
\beq
\omega_{*,s} = \frac{k_y \rho_s \vths{s}} {2L_{n,s}}\frac{q_s}{|q_s|},
\eeq
$\vths{s}=2T_s/m_s$, $m_s$ is species mass, $\rho_s = \vths{s}/|\Omega_s|$, and $1/L_{n,s}=-\partial \ln n_s/\partial x$.

From Poisson's Equation~(\ref{eqn:poisson}),
\beq
\int d^3 v  \left(h_i - h_e\right) = \frac{e n}{T_e}\left(1 + \tau - \lambda_e^2\nabla_{\perp}^2\right) \dphi,
\label{eqn:QN}
\eeq
where $\tau \doteq T_e/T_i$, and we have restricted our attention to a single ion species
with proton charge $e$.  Substitution of Eq.~(\ref{eqn:hks}) into Eq.~(\ref{eqn:QN}) results
in the dispersion relation
\beq
\epsilon(\omega,\mbf{k})=1+\tau + k_{\perp}^2\lambda_e^2 + \left(\zeta_e - \frac{k_y\rho_e}{2k_{\parallel} L_n }\right)
\modgam(k_y\rho_e) Z\left(\zeta_e\right)
+ \tau \left(\zeta_i + \frac{k_y{\rho_i}}{2k_{\parallel} L_n}\right)\modgam\left(k_y\rho_i\right)Z\left(\zeta_i\right) = 0,
\label{eqn:disp}
\eeq
where $\zeta_s \doteq \omega/k_{\parallel}\vths{s}$,
\beq
Z(x) \doteq \zi \sqrt{\upi} \textnormal{e}^{-x^2} \ \erfc{-\zi x}
\eeq
is the plasma dispersion function, $\textnormal{erfc}$ is the complementary error function, and
\beq
\modgam(x) \doteq \exp\left(-\frac{x^2}{2}\right) I_0\left(\frac{x^2}{2}\right),
\eeq
with $I_0$ a modified Bessel of the first kind.  Note that we have used quasineutrality to
set $L_{n,i}=L_{n,e}\doteq L_n$.

\subsection{Quasilinear energy exchange}

Using Eq.~(\ref{eqn:hks}) for $\hks$ in Eq.~(\ref{eqn:xchangek})
we get a quasilinear approximation for the energy exchange:
\beq
\begin{split}
\hat{H}_{\mbf{k},s} &\doteq \frac{\tilde{H}_{\mbf{k},s}}{n T_i} \frac{|L_n|}{\vths{i}}\frac{T_i^2}{e^2 |\tilde{\dphi}_{\mbf{k}}|^2} \\
&=\modgam(k_{\perp}\rho_s)\kpar |L_n|\frac{T_i}{T_s} \real{\zeta_i} 
 \imag{\left(\zeta_s + \frac{k_y\rho_s}{2\kpar L_n}\frac{q_s}{|q_s|}\right) Z(\zeta_s)},
\label{eqn:HQL}
\end{split}
\eeq
It is straightforward to verify that summing this expression over species and using the
dispersion relation of Eq.~(\ref{eqn:disp}) leads to zero net heating.  

This quasilinear estimate provides information about linear phase relationships that indicate the sign
of the heating contribution as a function of wavelength.  It predicts neither the spectrum nor the saturated heating amplitude in steady-state.  As we are primarily interested in the sign of the heating, it is 
enough to make some assumption about the turbulent heating spectrum; in particular, we assume that the steady-state heating has the same
sign as the quasilinear estimate for the mode(s) with the largest linear growth rate. 
In the following subsections we obtain numerical and approximate analytical solutions
for the mode frequencies and associated quasilinear heating in various limits.

\subsection{Comparable thermal speeds}

We first show that there is no instability, and thus no turbulent heating, if $\vths{e}=\vths{i}$.  To find the condition for marginal stability, we seek solutions for which $\gamma\doteq\imag{\omega}=0$.
In this case, the plasma dispersion function simplifies to $Z(x)=\sqrt{\upi} \exp(-x^2) (\zi - \erfi{x})$, with $\textnormal{erfi}$ the imaginary error function and $x=\real{\zeta}$.  The constraint $\imag{\epsilon}=0$ then gives
\beq
\left(x_e - \frac{k_y\rho_e}{2k_{\parallel} L_n }\right)
\modgam(k_y\rho_e) \exp(-x_e^2)=
-\tau \left(x_i + \frac{k_y{\rho_i}}{2k_{\parallel} L_n}\right)\modgam\left(k_y\rho_i\right)\exp(-x_i^2).
\eeq
Substituting this expression into the constraint $\real{\epsilon}=0$ gives
\beq
0 = 1+\tau +k_{\perp}^2\lambda_e^2+\tau \left(x_i + \frac{k_y{\rho_i}}{2k_{\parallel} L_n}\right)\modgam\left(k_y\rho_i\right)\upi^{1/2}\exp(-x_i^2) \left(\erfi{x_e}-\erfi{x_i}\right).
\label{eqn:ReEps}
\eeq
When $\vths{e} = \vths{i}$, then $x_e = x_i$, and Eq.~(\ref{eqn:ReEps}) has no solution.  Such a plasma is therefore either always stable or always unstable, independent of wavenumber and density gradient.  
As there is no instability for zero density gradient, the plasma must therefore be always stable for
$\vths{e}=\vths{i}$.

Next, we consider how marginal stability is modified when $\vths{e} = \vths{i}\left(1+\delta\right)$, with $|\delta| \ll 1$.  In this limit, the constraint Eq.~(\ref{eqn:ReEps}) becomes
\beq
0 \approx 1 +\frac{k_{\perp}^2\lambda_e^2}{2}- x_i \modgam(k_y\rho_i) \left(x_i + \frac{k_y\rho_i}{2\kpar L_n}\right)\delta,
\eeq
with solutions given by
\beq
x_i = -\frac{k_y\rho_i}{4\kpar L_n}\pm \sqrt{\left(\frac{k_y\rho_i}{4\kpar L_n}\right)^2
+ \frac{1+k_{\perp}^2\lambda_e^2/2}{\modgam(k_y\rho_i)\delta}}.
\label{eqn:xi}
\eeq
In order for the solutions to be real (as we assumed when we considered marginal stability),
the term inside the square root must be positive definite.  This constraint is satisfied when $\delta>0$ or when $(k_y\rho_i)^2 \modgam(k_y\rho_i) |\delta| > 16(\kpar L_n)^2(1+k_{\perp}^2\lambda_e^2/2)$.  The latter constraint can always be satisfied for sufficiently long parallel wavelengths.
Consequently, an unbounded system can be unstable for all finite values of $\delta$; i.e., for all plasmas
with $\vths{i}\neq \vths{e}$.

We now proceed to obtain the sign of the turbulent heating driven by instabilities with
$\delta <0$ and $\delta >0$, respectively.  Since $|\delta| \ll 1$, we 
can can approximate $\zeta_s \approx x_s$ in the quasilinear heating expression~(\ref{eqn:HQL}) to obtain:
\beq
\hat{H}_{\mbf{k},i} = \kpar |L_n| x_i
\modgam(k_{\perp}\rho_i) \left(x_i + \frac{k_y\rho_i}{2\kpar L_n}\right)\frac{\sqrt{\upi}}{2}\exp(-x_i^2).
\label{eqn:HQLapp}
\eeq
When $\delta < 0$, $x_i$ given by Eq.~(\ref{eqn:xi}) satisfies $-k_y \rho_i/2\kpar L_n < x_i < 0$.
Substituting this range of $x_i$ values in Eq.~(\ref{eqn:HQLapp}) results in the constraint
$\hat{H}_{\mbf{k},i} < 0$.  When $\delta > 0$, $x_i$ given by Eq.~(\ref{eqn:HQL})
satisfies $x_i > 0$ or $x_i < -k_y\rho_i / 2\kpar L_n$.  Substituting this range of $x_i$ values
in Eq.~(\ref{eqn:HQLapp}) results in the constraint $\hat{H}_{\mbf{k,}i} > 0$.
Combining these two constraints gives the general expression $\textnormal{sgn}(\hat{H}_{\mbf{k},i})=\textnormal{sgn}(\delta)$.  Thus we see that for small deviations from stability,
the quasilinear turbulent heating acts to equalize the ion and electron thermal speeds 
and thus stabilize the mode.

For the case of pair plasmas ($m_i = m_e$) our quasilinear analysis indicates that turbulent heating
driven by the electron drift wave acts to equalize the ion and electron temperatures.
We can also use a symmetry of the gyrokinetic-Poisson system of equations
to see how pair plasma heating depends on temperature ratio.  
In particular, we consider how the equations
are modified under an interchange of the electron and ion temperatures.  
First we note that the average over Larmor angle is unaffected by the interchange, as the Larmor radius
of each particle is independent of temperature.  Denoting the solutions when $\tau=\tau_0$ as
($h_s$, $\dphi$) and the solutions when $\tau = 1/\tau_0$ as ($\hswap{s}$, $\phiswap$), we have
\beq
\pd{\hswap{s}}{t} + \vpar\pd{\hswap{s}}{z} + \frac{c}{B}\{\left<\dphi^{\leftrightarrow}\right>_{\mathbf{R}},h_s^{\leftrightarrow}\} = \frac{q_s}{T_{s'}} \pd{\left<\phiswap\right>_{\mathbf{R}}}{t} F_{0,s'}+ \frac{c}{B}\pd{\left<\phiswap\right>_{\mathbf{R}}}{y}\pd{F_{0,s'}}{x} + C[\hswap{s}]
\eeq
and
\beq
\int d^3v \left(\hswap{i} - \hswap{e}\right) = \left(\sqrt{\tau_0} + \sqrt{\frac{1}{\tau_0}} - \frac{T_0}{4\pi e^2}\nabla^2\right) \frac{e\phiswap}{T_0},
\eeq
with $T_0 \defeq (T_e T_i)^{1/2}$, $s'=i$ when $s=e$, and $s'=e$ when $s=i$.  This admits 
solutions ($\hswap{s}(x,y,z,\vpar,v_{\perp},t) = -h_{s'}(-x,-y,z,\vpar,v_{\perp},t)$, $\phiswap(x,y,z,t)=\dphi(-x,-y,z,t)$).  The linear growth rates for pair plasmas are thus symmetric under interchange of $T_i$ and 
$T_e$, as seen in Fig.~(\ref{fig:gamma_pair}).

Applying these symmetries to the heating expression~(\ref{eqn:xchange0}) gives
\beq
\begin{split}
H_s^{\leftrightarrow} &= \frac{1}{V}\int d^3 R\int_{\mbf{R}} d^3 v q_s \overline{\left(\hswap{s}\pd{\gyroR{\phiswap}}{t}\right)} \\
&=\frac{1}{V}\int d^3 R\int_{\mbf{R}} d^3 v q_{s'} \overline{\left(h_{s'}(-x,-y,z,\vpar,v_{\perp},t)\pd{\left<\dphi(-x,-y,z,t)\right>_{\mathbf{R}_{s'}}}{t}\right)} \\
&= H_{s'} = -H_s,
\end{split}
\eeq
where the last equality follows from the fact that the species-summed heating is zero.  Thus
the heating of each species of a pair plasma is anti-symmetric under interchange of $T_i$ and $T_e$,
and there can be no turbulent heating when $T_i = T_e$.

\subsection{Disparate thermal speeds}

For plasmas with $m_i \gg m_e$, the ions and electrons will have
disparate thermal speeds $\vths{i} \ll \vths{e}$
for temperature ratios satisfying $m_e/m_i \ll \tau \ll m_i/m_e$.  In this case,
we can look for solutions to the dispersion relation that satisfy
$|\zeta_e| \ll 1 \ll |\zeta_i|$.  With this restriction, the plasma dispersion functions
appearing in Eq.~(\ref{eqn:disp}) can be greatly simplified.  In particular, we use
\beq
Z\left(\zeta_e\right) \approx \textnormal{i}\sqrt{\upi}
\eeq
and
\beq
Z(\zeta_i) \approx -1/\zeta_i,
\eeq
giving the following approximate dispersion relation:
\beq
\zeta_i\left(1+\tau + k_{\perp}^2\lambda_e^2\right) + \zeta_i\left(\zeta_e - \frac{k_y\rho_e}{2k_{\parallel} L_n }\right)
\modgam(k_y\rho_e) \iunit\sqrt{\upi}
- \tau \left(\zeta_i + \frac{k_y{\rho_i}}{2k_{\parallel} L_n}\right)\modgam\left(k_y\rho_i\right)= 0.
\label{eqn:disp_vte_gg_vti}
\eeq

\begin{figure}
  \centerline{\includegraphics[height=1.9in]{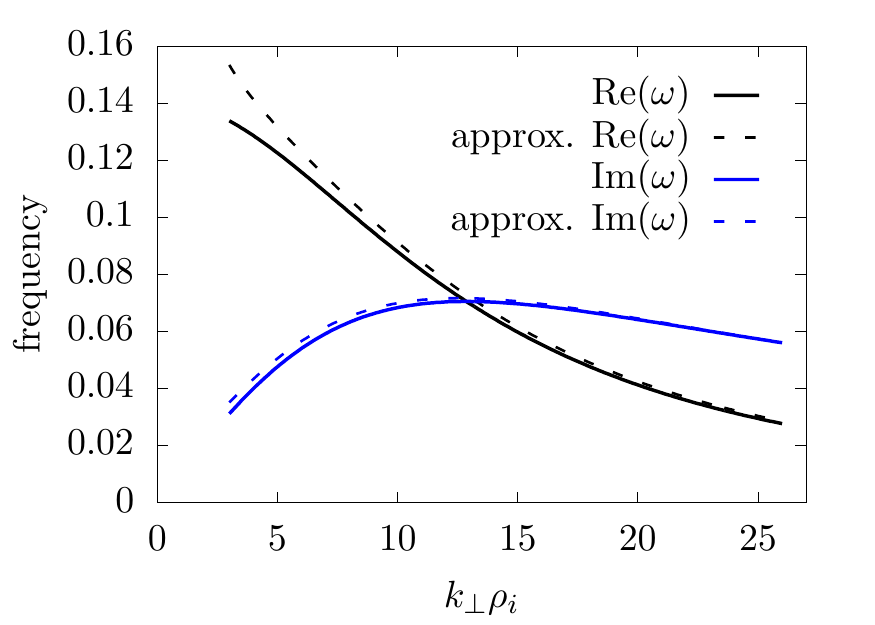}}
  \caption{Comparison of exact (solid lines) and approximate analytical (dashed lines) growth rates for the case $\tau=1$ and $m_i/m_e = 337824$ (tungsten ions).}
\label{fig:tungsten}
\end{figure}

We next consider wavelengths much shorter than the ion Larmor radius but much longer than
the electron Larmor radius; i.e., $k_{\perp}\rho_i \gg 1 \gg k_{\perp}\rho_e$.  In this limit
we can approximate $\Gamma(k_y\rho_e) \approx 1$ and $\Gamma(k_y\rho_i) \approx 1/(\sqrt{\upi}k_{\perp}\rho_i)$ in 
Eq~(\ref{eqn:disp_vte_gg_vti}) to obtain
\beq
\zeta_i\left(1+\tau + k_{\perp}^2 \lambda_e^2\right) + \zeta_i\left(\zeta_e - \frac{k_y\rho_e}{2k_{\parallel} L_n }\right) \iunit\sqrt{\upi}
- \frac{\tau}{\sqrt{\upi}} \left(\frac{\zeta_i}{k_{\perp}\rho_i} + \frac{k_y}{k_{\perp}}\frac{1}{2k_{\parallel} L_n}\right)= 0.
\eeq
Seeking solutions for which $\zeta_i \sim k_{\perp}\rho_e / \kpar L_n$ allows us
to neglect the $\zeta_e$ and $\zeta_i/k_{\perp}\rho_i$ terms.  The resultant solution
for $\omega$ is
\beq
\omega = \frac{\tau \kpar\vths{i}}{\sqrt{\upi}}\frac{2\kpar L_n\left(1+\tau+k_{\perp}^2\lambda_e^2\right)\textnormal{sgn}(k_y) + \zi \sqrt{\upi} k_{\perp}\rho_e}{\left(1+\tau+k_{\perp}^2\lambda_e^2\right)^2\left(2\kpar L_n\right)^2 + \upi k_{\perp}^2\rho_e^2}.
\label{eqn:omconventional}
\eeq
In Fig.~(\ref{fig:tungsten}) we show an example comparison of the exact and analytical expressions for $\omega$ for a plasma with $k_{\perp}\lambda_e=0$ and $m_i/m_e = 337824$ (corresponding to the
mass of tungsten).

Eq.~(\ref{eqn:omconventional}) indicates that for $\kpar > 0$ there is an instability with peak growth rate $\gamma$ at wavelengths $\mbf{k}_m$ satisfying $(\partial \gamma/\partial \kpar)|_{\mbf{k}_{m}} = (\partial \gamma/\partial k_{\perp})|_{\mbf{k}_m} = 0$.  These constraints give
\beq
\frac{2}{\sqrt{\upi}}k_{\parallel,m} |L_n| \left(1+\tau + k_{\perp,m}^2\lambda_e^2\right) = k_{\perp,m}\rho_e,
\eeq
with the largest growth rate for $k_{\perp,m}\lambda_e = 0$.  Using these results
in Eq.~(\ref{eqn:omconventional}) gives
\beq
\omega_m = \frac{\tau}{4\sqrt{\upi}\left(1+\tau\right)}\frac{\vths{i}}{|L_n|}
\left(\zi + \textnormal{sgn}(k_y L_n)\right),
\label{eqn:ommax}
\eeq
with $\omega_m\doteq \omega(\mbf{k}_m)$ the complex frequency evaluated at the wavevector $\mbf{k}_m$ that 
maximizes the linear growth rate.

Plugging Eq.~(\ref{eqn:ommax}) for $\omega_m$ into Eq.~(\ref{eqn:HQL}) for turbulent heating 
and using the appropriate approximations for $\modgam(k_y\rho_s)$ and $Z(\zeta_s)$ gives
\beq
\begin{split}
\hat{H}_{\mbf{k},i} = -\hat{H}_{\mbf{k},e} &\approx \frac{k_y\rho_e}{\kpar \vths{i}}\real{\omega}\frac{|L_n|}{L_n} \frac{\sqrt{\upi}}{4} \\
& = \frac{\tau}{8\sqrt{\upi}} > 0.
\end{split}
\eeq
So ions are heated and electrons are cooled; i.e., the instability acts to equalize 
$\vths{e}$ and $\vths{i}$ and thus stabilize the mode.

\section{Simulation results}

We now provide numerical data to verify our analytical predictions.  All simulations were conducted
using the local, Eulerian gyrokinetic code $\texttt{GS2}$~\citep{kotschCPC95,dorlandPRL00} 
with kinetic electrons and a single ion species
immersed in a straight, uniform magnetic field and with a cross-field density gradient.  The fluctuations
are constrained to be purely electrostatic, and we take $k_{\perp}\lambda_e = 0$.  Each simulation used 32 points along the magnetic field direction ($z$), and the velocity space is sampled on a polar grid~\citep{barnesPoP10a}, with 12 points in speed $v$ and 16 points in pitch angle $v_{\parallel}/v$.

\begin{figure}
\includegraphics[height=1.9in]{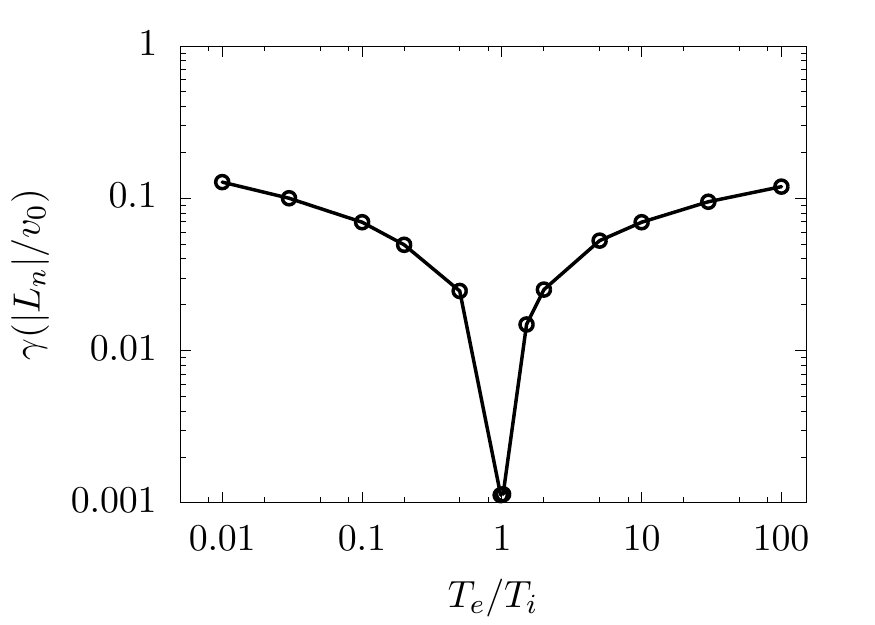}
\includegraphics[height=1.9in]{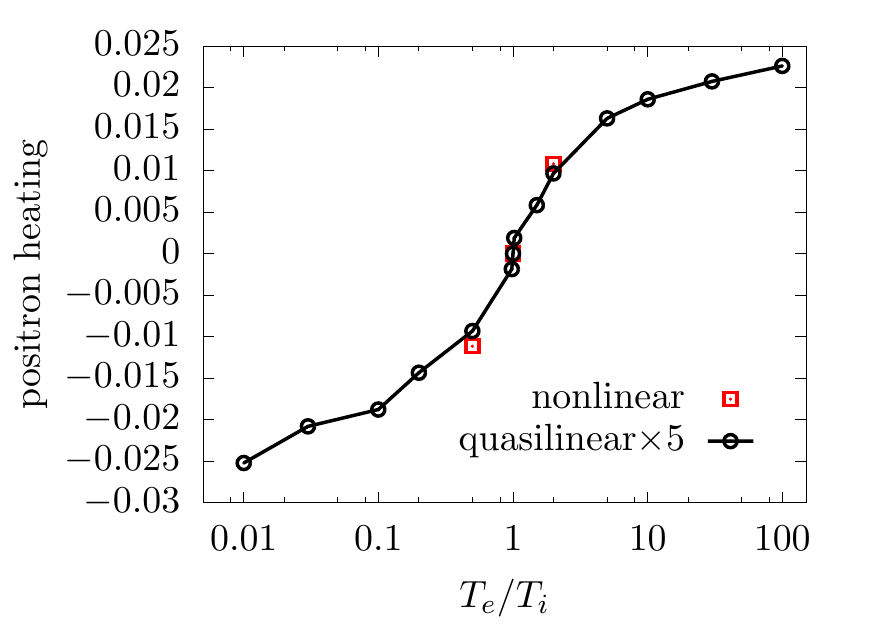}
  \caption{
  Normalized linear growth rates maximized over $\kpar$ and $k_y$ (left) and corresponding turbulent heating (right)
 from \texttt{GS2} simulations as a function of electron-ion temperature ratio for an electron-positron plasma.  The quasilinear turbulent ion heating $\hat{H}_i$ (black circles and line) given by Eq.~(\ref{eqn:HQL}) is weighted by the linear growth rate $\gamma |L_n|/v_0$ to qualitatively estimate saturated fluctuation amplitudes.  The turbulent ion heating $(H_i |L_n| / n T_0 v_0) (|L_n|/\rho_0)^2$
obtained from nonlinear simulations is also given (red squares).  Here $\rho_0 = v_0 / \Omega_i$, with
 $v_0$ the geometric mean of the ion and electron thermal speeds.
 }
\label{fig:gamma_pair}
\end{figure}

First we consider the case of an electron-positron plasma.  The linear growth rates, maximized
over $\kpar$ and $k_{\perp}$, are plotted against $T_e/T_i$ in Fig.~(\ref{fig:gamma_pair}).  They
are normalized by $|L_n|/v_0$, with $v_0 \defeq \sqrt{\vths{i}\vths{e}}$, in order to make manifest the expected symmetry of the growth rates with respect to the interchange of $T_i$ and $T_e$.  As predicted,
the plasma is stable only when $T_e = T_i$, and the growth rates are symmetric about $T_e/T_i=1$.  

Also shown in Fig.~(\ref{fig:gamma_pair}) are the quasilinear heating estimates defined in Eq.~(\ref{eqn:HQL}) and the heating determined from nonlinear simulations as a function of $T_e/T_i$.
The nonlinear simulations employed 47 modes (after de-aliasing) in both directions perpendicular to the magnetic field ($x$ and $y$).  As we showed analytically, the heating (both quasilinear and nonlinear) 
is antisymmetric about $T_e/T_i=1$, with the heating acting to equilibrate the electron and positron
temperatures and thus shut off the linear instability.

\begin{figure}
\includegraphics[height=1.9in]{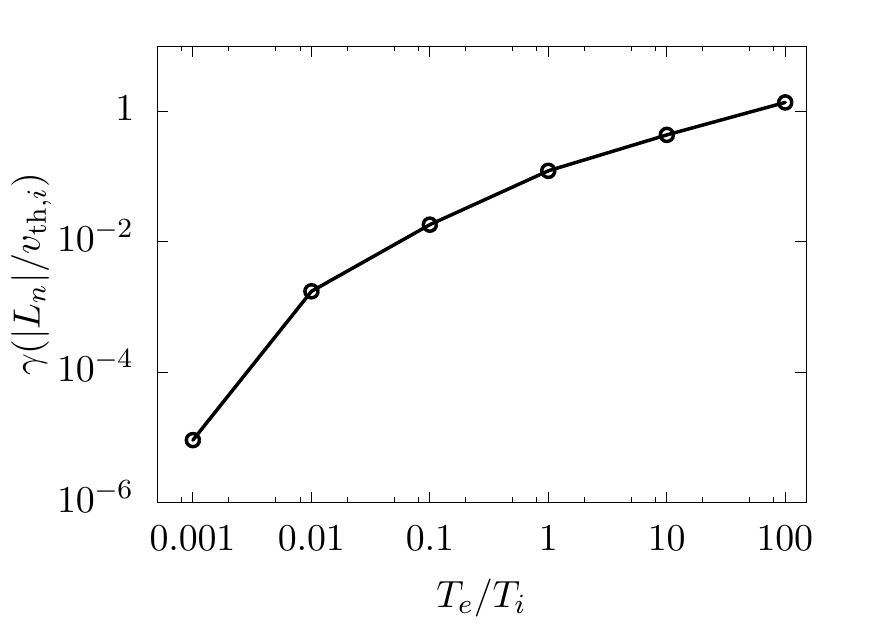}
\includegraphics[height=1.9in]{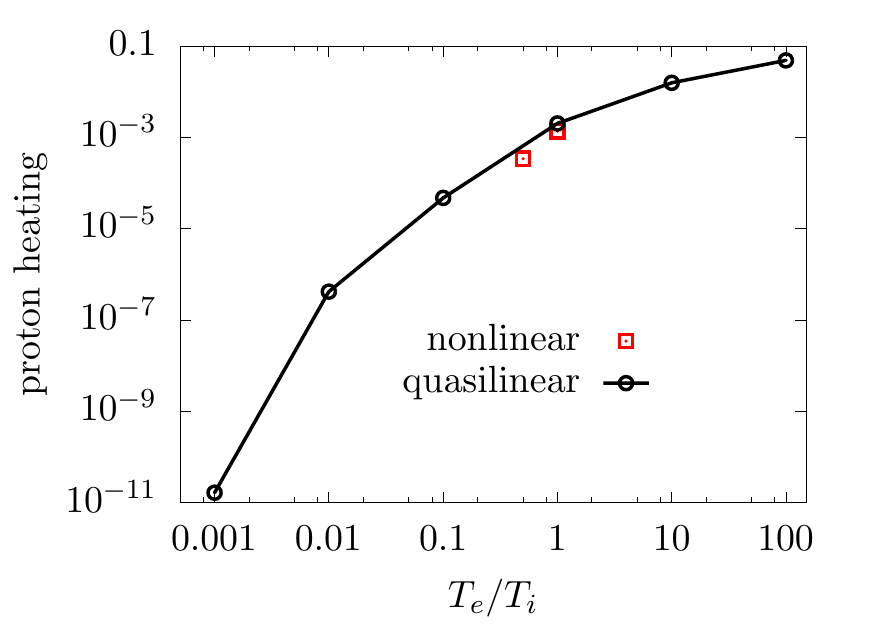}
  \caption{Normalized linear growth rates maximized over $\kpar$ and $k_y$ (left) and corresponding turbulent heating (right)
 from \texttt{GS2} simulations as a function of electron-ion temperature ratio for an electron-proton plasma.  The quasilinear turbulent ion heating $\hat{H}_i$ (black circles and line) given by Eq.~(\ref{eqn:HQL}) is weighted by the linear growth rate $\gamma |L_n|/\vths{i}$ to qualitatively estimate saturated fluctuation amplitudes.  The turbulent ion heating $(H_i |L_n| / n T_i \vths{i}) (|L_n|/\rho_i)^2$
obtained from nonlinear simulations is also given (red squares)}
\label{fig:gamma_proton}
\end{figure}

We next show the growth rates and turbulent ion heating as a function of electron-ion temperature
for an electron-proton plasma in
Fig.~(\ref{fig:gamma_proton}).  Again we see that the plasma is unstable for $\vths{e}\neq \vths{i}$.
Furthermore, the ion turbulent heating is positive definite for $\vths{e} > \vths{i}$, in agreement 
with our approximate analytic result.

\section{Conclusions}

The analytical and numerical results shown in this paper indicate that turbulent heating
driven by the electron drift wave instability present in an inhomogeneous, 
magnetized plasma acts to stabilize the mode.  Stabilization occurs when the ion and
electron thermal speeds are equal.  For a conventional mass ratio
plasma with $T_i \sim T_e$, this leads to the ions being heated and the electrons cooled; 
for a pair plasma, the turbulent heating acts to equalize the ion and electron temperatures.

While turbulent heating acts to stabilize the mode, it is not the only stabilization mechanism
in the system.  We showed that the instability drives particle transport that flattens the driving
density gradient on a time scale at least as fast as the heating influences the electron-ion temperature
ratio.  Consequently, for the turbulent heating to fully set the thermal equilibrium there must be additional
physics that fixes the density gradient; e.g., an external density source or a
lowest order equilibrium set by gravitational forces.
In this case, thermal equilibrium would correspond to $T_i /T_e = \sqrt{m_i/m_e}$.

The authors would like to thank F. I. Parra, A. A. Schekochihin and A. Zocco for useful discussions.  M. Barnes was supported in part by STFC grant ST/N000919/1.  The authors also acknowledge the use of ARCHER through the Plasma HEC Consortium EPRSC grant number EP/L000237/1 under project e281-gs2 and the use of 
the EUROfusion High Performance Computer (Marconi-Fusion) under project MULTEI.

\bibliographystyle{jpp}

\bibliography{general}

\end{document}